\documentclass{skaox2006}
\usepackage{graphicx}

\newcommand{\hompc}{\,h\,{\rm Mpc}^{-1}}

\begin{document}
   \title{Measuring Dark Energy with the Wide--Field Multi--Object Spectrograph (WFMOS)\thanks{On behalf of the WFMOS feasibility study team of Barden et al. (2005)}}

   \author{
          Robert Nichol\inst{1}
          }

   \institute{Institute of Cosmology and Gravitation (ICG), Univ. of Portsmouth, Portsmouth, PO1 2EG}

   \abstract{Dark energy is one of the greatest scientific challenges
of the $21^{st}$ century. One of the key questions facing cosmologists is
whether dark energy is either a breakdown of General Relativity on
large scales or a new form of matter in the Universe with a negative
effective pressure. This question can only be answered through a suite
of different observations as a function of redshift. In this paper, I briefly review various dark energy reports published in the last year, which all highlight the importance of the baryon acoustic oscillations (BAO) for probing the ``dark physics" of the Universe. I also summarize the recent measurements of the BAO in large galaxy redshift surveys. I then look forward to a new instrument planned by the Subaru and Gemini communities called the ``Wide--Field Multi--Object Spectrograph" (WFMOS) for the Subaru telescope. The baseline design of this facility includes $\simeq4500$ spectroscopic fibers over a field--of--view of 1.5 degree diameter, covering a wavelength range of $0.39$ to 1 microns. The instrument is schedule for first--light early next decade and will perform massive spectroscopic surveys of both distant galaxies and faint stars in our own Galaxy.  The WFMOS dark energy surveys will deliver $\simeq1$\% errors on the angular--diameter distance and Hubble parameter to high redshift. WFMOS will also be a unique user--facility allowing astronomers to address a host of astrophysical problems like galaxy evolution, the intergalactic medium and calibrate photometric redshifts. The WFMOS archive will also provide a rich resource for further ancillary science much like the present--day SDSS archive.

}
   \maketitle
%
%
\section{Introduction}

The most striking discovery in astrophysics over the last ten years is
that the energy density of the Universe is dominated by a mysterious
quantity called ``dark energy'' (Spergel et al. 2006). This dark
energy is responsible for a late-time acceleration of the Hubble
expansion of the Universe and its exact nature remains unclear. In
its simpliest form, dark energy could be Einstein's Cosmological
Constant ($\Lambda$), yet its observed value is substantially smaller
than expected for the vaccum energy density. Dark energy could be another scalar field that evolves with cosmic time, e.g.,
Quintessence. Alternatively, the late-time acceleration of the Universe could be the result of our lack of understanding of gravity
on large--scales, and many authors have recently proposed modified
gravity models to account for these cosmological observations
(e.g. Fairbairn \& Goobar 2005; Maartens \& Majerotto 2006).

The next decade will be dominated by new efforts to measure dark
energy to greater precision and therefore, determine its true nature. In particular, we can attempt to answer two fundamental questions about dark energy:

\begin{enumerate}

\item {\it Is dark energy just the Cosmological Constant?} This
question will be addressed through accurate measurements of the
equation of state of dark energy ($p = w \rho \, c^2$) as a function of cosmological
time. A cosmological constant is given by $w=-1$, while quintessence models usually predict values in the range of $-1<w\le 0$.

\item {\it Is dark energy a modification of gravity, or a new form of matter?} This question will likely be addressed through probing the Universe using different methods and tracers of the dark energy? In particular, we may expect differences in the rate of growth of cosmic structures in the Universe (see Ishak, Upadhye \& Spergel 2006; Linder 2006; Huterer \& Linder 2006)

\end{enumerate}

These issues have been extensively discussed in a series of recent dark energy reviews by both national and international organizations. In particular, I highlight below three outstanding reviews of the dark energy physics and summarise their key recommendations.

\begin{itemize}

\item{{\it The Dark Energy Review} by Trotta \& Bower (astro--ph/0607066) which was commissioned by the Science Committee of PPARC. This report clearly favors weak lensing and BAO experiments to understand dark energy (see below). This is primarily due to their statistical accuracy and robustness to systematic uncertainties. They also highlight that WFMOS could be ``pioneering in the field of wide and deep spectroscopic reconstruction of the acoustic peaks" }

\item{{\it The Dark Energy Task Force (DETF)} by Albrecht et al. (astro-ph/0609591) in the US. This is the most comprehensive of the reports providing a quantitative ``figure of merit" for the various dark energy experiments and techniques. Their major recommendations include the use of multiple techniques to study dark energy, with at least one of these techniques being a probe of the  growth of structure in the Universe. They recommend immediate funding for projects that improved our understanding of systematics in the dark energy measurements, as these are now the dominant source of uncertainty.}

\item{{\it The ESA-ESO Working Group on Fundamental Physics} report by Peacock et al. The report focuses on european projects that could make significant  progress in understanding dark energy. The report recommends the undertaking of a space-borne imaging survey over a major fraction of the sky, complemented by photometric redshifts from new optical and infrared ground--based surveys. The report also highlights the importance of new spectroscopic surveys of $>10^5$ galaxies to calibrate the photometric redshifts and a new wide-field instrument for such calibrations maybe required. The report also notes the importance of improved local samples of supernovae to fully exploit them as cosmological probes. }

\end{itemize}

In general, all these reports highlight the 
need for new massive surveys of the Universe using dedicated
facilities. This is simply due to the fact that dark energy is a small
observational signal and thus requires big surveys to beat the
statistical noise (cosmic variance and shot--noise) and new
experiments to control the systematic uncertainties. Today, weak lensing (WL) and cluster
surveys are uncompetitive (compared to the existing BAO and SNe
surveys) but all the reports highlight that this will change in the coming
years, with WL surveys having the most promise. 

This situation is clearly detailed in the DETF report which
tabulates the fractional decrease in the error ellipse on the equation of state of dark energy as
a function of experiment type (BAO, SNe, clusters, WL). In particular,
they compare the error ellipses from the expected
pre--2010 experiments (called Stage I \& II experiments in their report, and
represent surveys already funded and underway), to the post-2010
experiments (Stage III \& IV), that are now being considered for funding around
the world. They also provide optimistic and pessimistic estimates for
the fractional gain in the errors to encompass our present understanding of the systematic uncertainties in each of these techniques.

Generally, the BAO and SNe methods have close optimistic and pessimistic
estimates indicating that these are now mature techniques with understood systematics. In constrast, the optimistic and pessimistic errors for WL
and clusters are widely discrepant reflecting the uncertainty with
these methods. However, the potential for large fractional increases
in our knowledge (in the optimistic case) is greatest for these two
techniques. In other words, WL and clusters offer the {\it ``high risk,
high gain''} options, while BAO and SNe are the {\it ``safe''} options
(although these techniques  will still deliver $>100$\% improvement in the errors on the
dark energy parameters after 2010).

   \begin{figure*}[tp]
   \centering
   \includegraphics[width=4.5in]{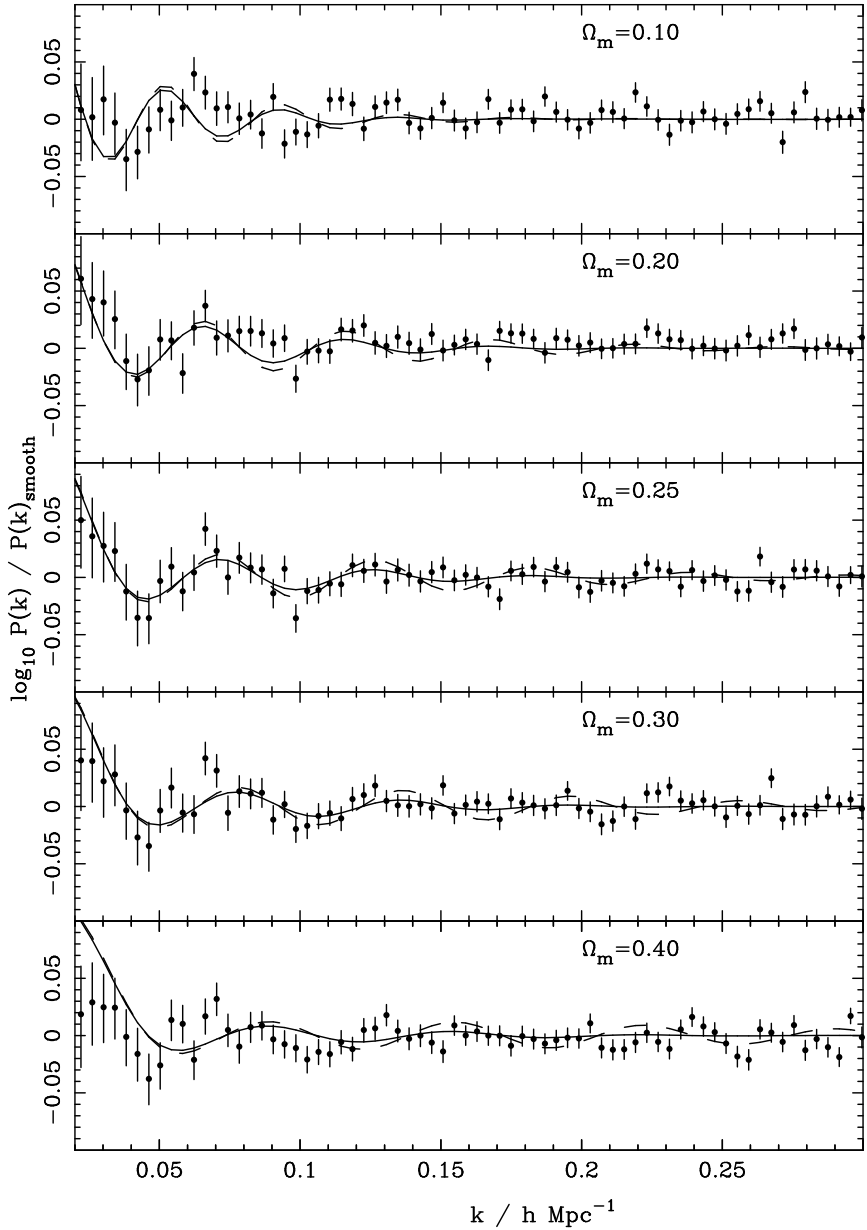}
   \caption{Figure taken from Percival et al. (2006a). This figure shows the ratio of the power spectra calculated from the SDSS DR5 dataset to
    a smooth cubic spline fit used to model the overall shape
    of the measured power spectra. This cubic spline was calculated using 8 nodes
    separated by $\Delta k=0.05\hompc$ and $0.025\le k\le0.375\hompc$,
    and an additional node at $k=0.001\hompc$. The data are plotted as solid circles (with 1 sigma errors) using five flat $\Lambda$
    cosmological models to convert from redshift to comoving distance,
    with matter densities given in each panel. That is why the data points change between the five panels. For comparison, in each
    panel we also plot the BAO predicted by a CDM model with the same
    matter density, $h=0.73$, and a 17\% baryon fraction (solid
    lines). The dashed lines show the same models without the
    low-redshift small-scale damping term. As can be seen, the
    observed oscillations approximately match those predicted by this
    model for $0.2\le\Omega_M\le0.3$, but fail for higher or lower
    matter densities.)
            \label{fig:percival}
           }
    \end{figure*}

I note here that all these reports discount the role of the late-time  Integrated Sashs-Wolfe (ISW) effect in constraining dark energy due to its low statistical power (caused by cosmic variance). I would however stress the importance of the ISW as it directly probes the effect of dark energy on the growth of cosmic structures in the Universe and can be measured to high redshift ($z>1$), thus testing the dark energy paradigm in a unique and complementary way (see Giannantonio et al. 2006). The ISW effect provides an ``insurance policy" against missing rapid changes in $w(z)$, at high redshift. Pogosian et al. (2005)  demonstrated this as ``the cross-correlation of Planck CMB data and LSST galaxy catalogs will provide competitive constraints on $w(z)$, compared to a SNAP-like SNe project, for models of dark energy with a rapidly changing equation of state".

\begin{table*}
\centering
{\large
\caption{The baseline design of WFMOS on Subaru taken from the feasibility study of Barden et al. (2005) and the white paper of Glazebrook et al. (2005)}
\begin{tabular}{|l|l|}\hline
Field--of--view & 1.5 degree diameter\\ \hline
Wavelength range & $0.39$ to $1.0$ microns\\ \hline
Low resolution spectrographs & 10 with similar design as SDSS\\
					&  (R=1800 in blue, R=3500 in red)\\ \hline
High resolution spectrograph & R=40000 for galaxy archeology science\\ \hline
Fibers & 3000 for low resolution (1 arcsec aperture)\\ 
            & 1500 for high resolution \\ 
            & ``Echidna"  fiber-positioner at prime focus (like FMOS on Subaru)\\ \hline
 Data analysis & Reduced ``on-the-fly" \\
 			& ``Nod \& Shuffle" for optimal sky subtraction\\
			& Data archived immediately for ancillary science\\ \hline
			
\end{tabular}
}
\end{table*}

In summary, which of these techniques should the astronomical community pursue? The
answer is the same advice one would receive when considering a pension
funds; {\bf Diversify!} Some of our assumptions about the
systematic errors will likely be wrong, so we need ``safe" options to
spread this risk. Meanwhile, the riskier options provide orthogonal
information and therefore, will lead to a greater understanding beyond
the simple sum of the parts.

\section{Baryon Acosutic Oscillations}

The baryon acoustic oscillations (BAO) have received considerable
attention over the last 5 years and have emerged as a key technique
for measuring $w(z)$ as outlined in the reports above. The BAO are
caused by sound waves propogating through the primodial plasma in the
early Universe. At recombination, these sound waves are frozen into
the distribution of matter as a prefered scale given by the
$\simeq0.57ct$, where $c$ is the speed of light and $t$ is the age of
the Universe since the Big Bang. Therefore, the BAO represent a
standard ruler in the Universe, which is left imprinted in the
distribution of matter. See Eisenstein \& Hu (1998) for a
comprehensive review of the physics of the BAO, or Bassett, Nichol \& Eisenstein (2005) for a popular review of the BAO.

The BAO standard ruler has already been measured at the surface of last scattering
as the Doppler peaks in the CMB power spectrum. This
provides an accurate estimation for the distance to this surface (Spergel
et al. 2006). Clearly, if one can detect and measure the BAO at other
redshifts, then one can jointly constrain the geometry of the Universe
and its content as a function of redshift.

In the last 5 years, there have been several measurements of the BAO in the
distribution of galaxies in the late Universe. In 2001, Miller, Nichol \& Batuski (2001) and Percival et al. (2001) presented first evidence for the
BAO in the Abell cluster catalogue and 2dFGRS respectively. For example, Miller et
al. (2001) obtained constraints on the cosmological parameters that are fully consistent with the present--day best--fit
cosmology (Spergel et al. 2006; Percival et al. 2006b; Tegmark et
al. 2006). In 2005, both the SDSS and 2dFGRS provided convincing
evidence for the BAO in the distribution of local galaxies (see
Eisenstein et al. 2005; Cole et al. 2005). In 2006, Percival et al. (2006a)
presented a detailed analysis of the SDSS DR5 galaxy redshift survey
and provides a $3\sigma$ detection of the BAO signal independent of the shape
of the power spectrum. This new SDSS
measurement is shown in Figure 1. Furthermore, Percival et
al. (2006a) used the BAO scale to determine
$\Omega_m=0.256^{+0.029}_{-0.024}$ (a $\simeq$10\% measurement), which again is
independent of the shape of the power spectrum and thus, independent
of concerns about scale--dependent biasing (assuming it is a smooth function of scale). 

Therefore, the detection of the BAO in the local galaxy distribution
is clear (Figure 1) and recent work demonstrates that it can
deliver robust and competitive measurements of the cosmological
parameters. It is interesting to note that the $\Omega_m$ value
derived from the BAO scale alone is in excellent agreement with
the value of $\Omega_m$ derived from the overall shape of the power spectrum,
i.e., the horizon scale ($\Omega_m h^2$) from the turn--over in the
power spectrum on large scales. Percival et al. (2006b) finds
$\Omega_m=0.22+/-0.04$ from the analysis of the shape of the SDSS DR5 power spectrum,
which includes accurate modeling of the luminosity--dependent biasing
of galaxies. 

\section{WFMOS Instrument}
 \begin{figure*}[tp]
   \centering
   \includegraphics[width=5in]{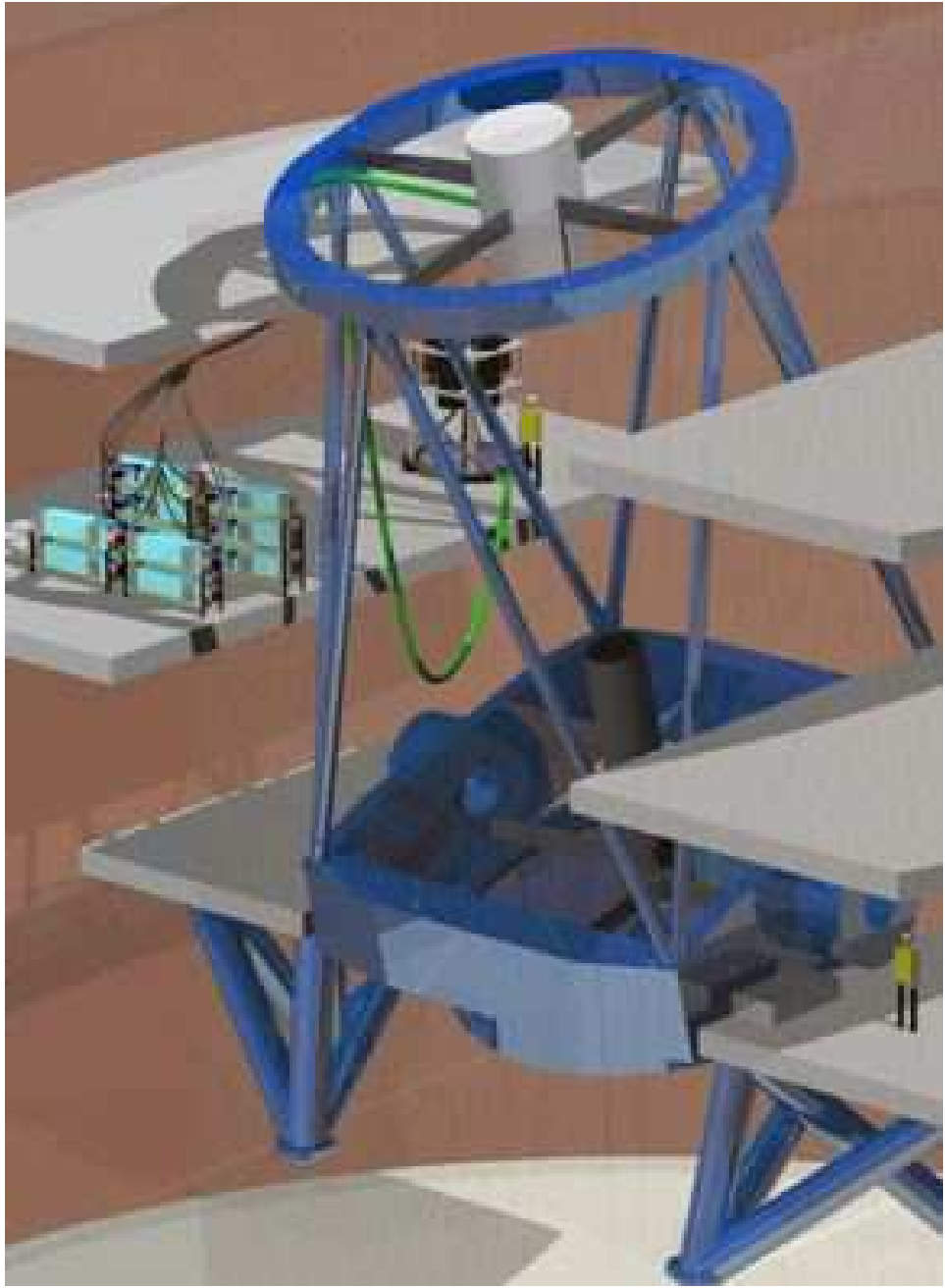}
   \caption{A CAD rendering of WFMOS on Subaru. The prime focus echidna fiber-positioner feeds $\simeq10$ SDSS-like low--resolution spectrographs and a high--resolution spectrograph located in a special room close to the telescope.              \label{subaru}
           }
    \end{figure*}

The proposed Wide-Field Multi-Object Spectrograph (WFMOS) started life as the KAOS concept instrument (see http://www.noao.edu/kaos/) and emerged as one of the key future instruments of the Gemini Observatory. This decision was made via the ``Aspen Process", which was a community--based discussion to determine the key scientific questions for Gemini to address in the future, and the types of instruments that might be designed and built to address them. Based on all this input, the Gemini Science Committee (GSC) recommended three top-priority instruments, which included WFMOS . In November 2003, the Gemini Board endorsed the GSC decision and initiated a feasibility study of the WFMOS instrument on both the Gemini and Subaru telescopes. 

The WFMOS feasibility study was delivered in February 2005 and was fully reviewed by Gemini and Subaru. Both the GSC and Gemini Board agreed that this study proved the feasibility of WFMOS on Subaru and {\it ``WFMOS offers the most transformative science opportunities"$\footnote{Quote taken from the Gemini website concerning the ``Aspen process"}$}. Gemini is now engaged in two competing design studies to determine the detailed technical viability and cost of WFMOS. These studies will hopefully deliver reports to Gemini and Subaru in late 2007, allowing the construction of WFMOS to begin in earnest in 2008. 

The baseline design of WFMOS has been presented in the WFMOS feasibility study of Barden et al. (2005) and is summarized in the white paper to the US DETF of Glazebrook et al. (2005). We present an overview of this baseline design in Table 1 and a CAD rendering of WFMOS on the Subaru telescope in Figure 2.  

\section{WFMOS Surveys}

\begin{table}
\caption{The baseline DE surveys of WFMOS taken from the feasibility study of Barden et al. (2005) and the white paper of Glazebrook et al. (2005)}
\centering
\begin{tabular}{|c|c|c|c|c|c|}\hline
z range & $r_{AB}$ & Vol. & Area & No. of & Nights \\ 
               &                  & (Gpc$^3$) & (deg$^2$) & galaxies & \\ \hline
0.5 -- 1.3 & 22.7& 4 & 2000 & $2\times10^{6}$ & $\simeq$100 \\ \hline
2.3 -- 3.3 & 24.5& 1 & 300 & $6\times10^{5}$ & $\simeq$100 \\ \hline
\end{tabular}
\end{table}

The primary goal of WFMOS is to facilitate efficient mapping of the night sky and therefore, three major astronomical surveys are envisaged for WFMOS to address the nature of dark energy, through precision measurements of its equation of state, and understanding the formation of our Galaxy, via detailed chemical and dynamical studies of stars in the Galaxy. 

For the DE study, two baseline galaxy redshift  surveys are planned as outlined in Table 2. The first will be a massive survey of $z\sim1$ emission--line galaxies covering 2000deg$^2$ of sky, while the second will target $z\sim3$ Lyman--break galaxies over several hundred square degrees. In this way, WFMOS will measure the BAO scale within two complementary redshift  shells, thus providing two accurate measurements of the angular diameter distance at lookback times of $\simeq8$Gyrs and $\simeq12$Gyrs\footnote{Assuming a flat cosmology with $h=0.7$ and $\Omega_m=0.25$} respectively (or approximately 60\% and 85\% the age of the Universe). Combined with the SDSS BAO measurements at $z<0.4$ (Percival et al. 2006a), and other planned surveys like ``WiggleZ" (at $z\sim0.7$), we will be able to map the angular--diameter distance with redshift over most of the observable Universe.

 \begin{figure*}[tp]
   \centering
   \includegraphics[width=5in]{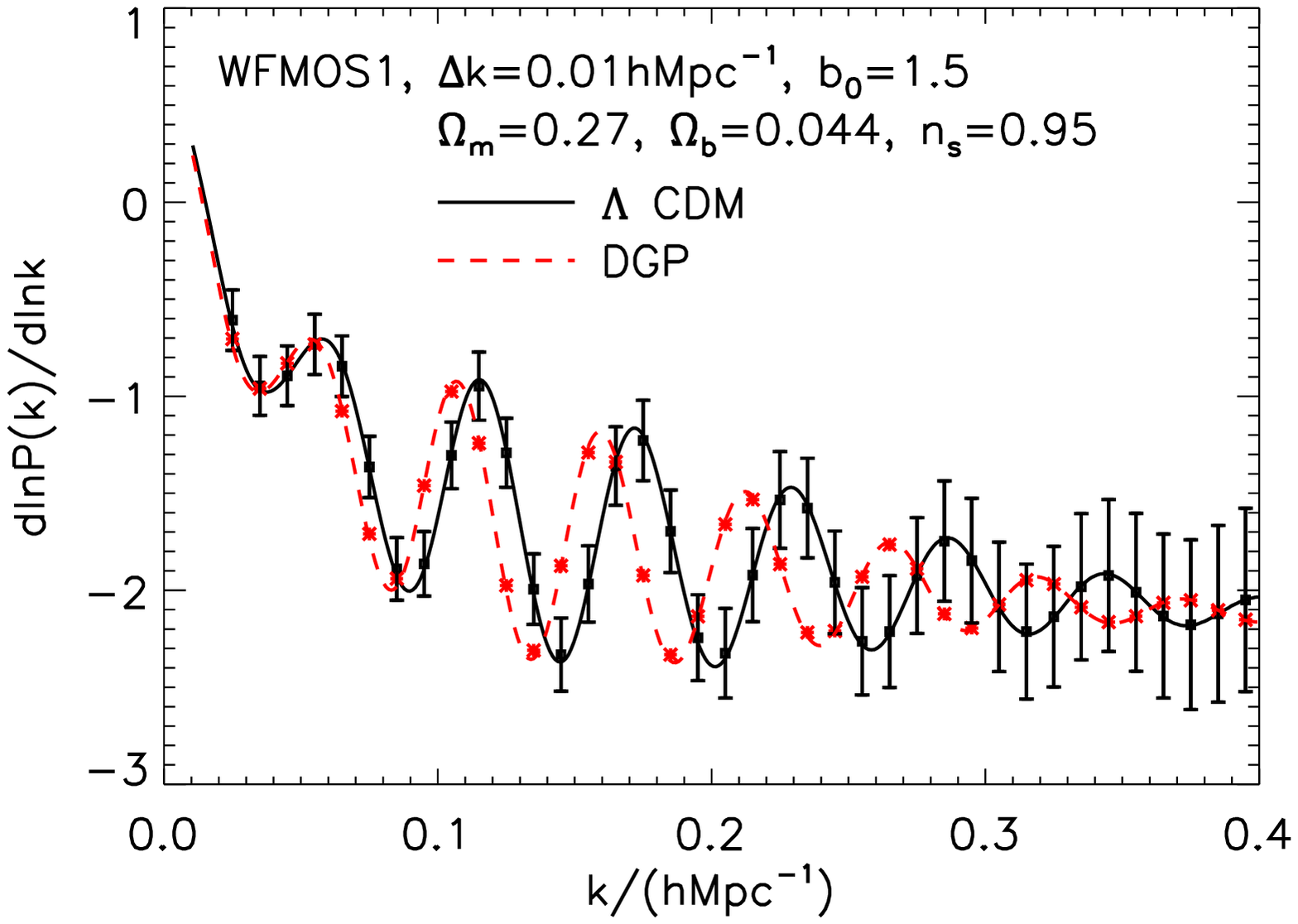}
   \caption{Figure 9 of Yamamoto et al. (2006).  The figure shows theoretical predictions of $d\ln P/d\ln k$ for the WFMOS $z\simeq1$ sample based on the $\Lambda$CDM (solid black line) and
modified DGP model (dashed red line). The data
points with error bars are derived from mock data with the
wavenumber binsize of $\Delta k=0.01 h{\rm Mpc}^{-1}$.  Based on chi--square, Yamamoto et al. find that these
two models are different by 6 $\sigma$.    \label{dgp}
           }
    \end{figure*}


Glazebrook et al. (2005) presented estimates for the precision of WFMOS measurements of $w(z)$ given the baseline surveys in Table 2.  First, we expect these surveys to deliver an approximate $1\%$ error measurement of the angular diameter distance ($D_A(z)$) and Hubble parameter ($H(z)$) at both $z\simeq1$ and $z\simeq3$. These cosmological measurements translate to a 5\% error ($1\sigma$) in $w_0$ and a 20\% error in $dw/dz$, when combined with Planck CMB and ground-based SNe data. This is as good a constraint on $w(z)$ as expected from future space--based SNe experiments (see Glazebrook et al. 2005). 

It is worth stressing that these estimates on the error on $w_0$ and  $dw/dz$ are conservative because they assume a near-flat underlying fiducial cosmological model (close to $w=-1$) when computing these errors. Such flat models disfavor dark energy observations at high redshift as the density of dark energy decreases with redshift. This point is discussed by Bernstein (2006) where the combination of the next generation of BAO and weak lensing measurements can provide tight constraints on the underlying geometry of the Universe, independent of the form of $w(z)$.  It would be a shame to miss a slightly non--flat cosmology because we thought $w(z)$ was exactly $-1$

The dark energy constraints presented above are now being revisited using the methods outlined in Bassett, Parkinson \& Nichol (2005) and Bassett (2005). This methodology also allows for the optimization of WFMOS surveys over a range of possible dark energy models and thus different parameterization of $w(z)$. See also the work by Blake et al. (2005) and Glazebrook \& Blake (2005).

Finally, WFMOS also provides an opportunity to test models of modified gravity. As shown in Figure 3, taken from Yamamoto et al. (2006), the WFMOS BAO measurements can be used to differentiate between the DGP model, which allows gravity to leak into extra dimensions on large scales (thus decreasing the strength of gravity), and the standard $\Lambda$CDM model. These predictions reflect the effect of the relative change in 
the background geometry between these two models. The growth of cosmic 
structure is different in these models because the Poisson equation 
is modified in the DGP model, but this effect is small in the figure. 
In order to measure the difference in the growth of structure, higher order correlation 
statistics and the weak lens measurement might be useful.

\section{WFMOS Facility}


In addition to the dedicated surveys discussed above, it is envisaged that WFMOS will be a powerful facility instrument for the Gemini and Subaru communities. This will likely be a  proposal--based allocation of time for specific science projects by small teams of astronomers. Many examples of such PI-driven science were outlined in the original KAOS proposal (www.noao.edu/kaos) and reproduced in the appendix to the WFMOS feasibility study. I highlight below a few of these (smaller) projects that could be achieved with a few nights on WFMOS:

\begin{itemize}

\item{\it Obtaining spectra for all galaxies in the Coma Cluster.} There are approximately 100,000 galaxies per deg$^2$ (Bernstein et al. 1995; Milne et al. 2006) in the core of the Coma cluster ($z=0.0232$) to an absolute magnitude limit of $M_R=-11$ (which corresponds to luminous globular clusters). This limit is $r=24$, which should be achievable with WFMOS and therefore, one could map (in a reasonable time) the total galactic content of this nearby massive cluster.

\item{\it Calibrate photometric redshifts} As recently outlined in the Peacock et al. report (see Introduction), it will be imperative to accurately calibrate photometric redshifts for the next generation of dark energy imaging surveys (DES, LSST, etc.). They estimate that this will require $>10^5$ redshifts selected in an unbiased way. It may also be important to obtain these photo--z calibration over the entire area of the imaging survey thus minimizing the effect of photometric zero--point uncertainties  across the survey. Such calibrations could be achieved with WFMOS and combined with an unbiased spectroscopic study of faint galaxies. 

\item{\it Probe the ISM using densely sampled quasars and galaxies.} Using WFMOS, one could simultaneously observe QSOs and galaxies in the same fields, thus studying the correlations between the Lyman-alpha forest and the intervening galaxy population. 

\end{itemize}

\section{WFMOS Archive}

Like the SDSS and 2dFGRS, any massive redshift survey will generate a host of ancillary results based on archival data. The plan for WFMOS is to reduce the spectra ``on--the--fly" and place both the reduced spectra, and derived science quantities (equivalent width, redshifts, etc.), straight into the Gemini Science Archive (GSA) as well as distributing it to member nations and involved scientists (the PI-driven projects will likely have a proprietary period before data is released to the public). Examples of archival science that would be possible include:

\begin{itemize}

\item{\it The serendipitous detection of high--redshift SN Ia's. } We will detect by accident nearly one thousand SN Ia's in the spectra of $z\sim 1$ galaxies (see Madgwick et al. 2003). Using the SDSS SNAFU project as reference, approximately a third of these Ia's will be detected before peak in their light curve and therefore, could be followed--up using imaging telescopes. This would provide a complementary method of detecting high redshift SNe with different systematic uncertainties compared to more traditional methods; for example, the spectroscopically--detected Ia's will be close to the centers of the host galaxies, while traditional photometric SNe searches typically avoid such SNe because of severe host galaxy contamination. 

\item{\it Testing General Relativity.} By comparing the angular-diameter distance to the luminosity-distance, one can test the Reciprocity relation of $d_L(z)=(1+z)^2\,d_A(z)$, which holds for any metric theory of gravity. One can also test dark matter--photon interactions, or any mechanism that causes the loss of photons (see Bassett \& Kunz 2004 for details) 

\item{\it The Alcock-Paczsynki effect.} See Matsubara (2004) and Yamamoto, Bassett \& Nishioka (2005).

\end{itemize}


\section{Conclusions}

As outlined in this paper, dark energy is a challenging problem for the whole of physics and will require significant resources and time to address its nature. Massive surveys of the Cosmos are required produced by new instruments that have tight control over their systematic errors. WFMOS is one such experiment which is planned to deliver the next generation of galaxy redshift surveys at high redshift. Predictions indicate that WFMOS will measure the angular diameter distance and Hubble parameter to 1\% accuracy, as well as potentially differentiate between models of modified gravity and $\Lambda$CDM. When combined with the planned HyperSuprimeCam on Subaru, these instruments together should become the next SDSS--like project, but on an 8--meter class telescope.

\begin{acknowledgements}
Many thanks to the organisers for the invite and their patience waiting for this manuscript. Many thanks to my collaborators for allowing me to show their work at the conference and herein. Thanks to the WFMOS feasibility study team  for their help and advice.

\end{acknowledgements}

\end{document}